\documentclass[12pt]{article}
\usepackage{amsxtra,amssymb,amsthm,amsmath,latexsym}

\theoremstyle{plain}

\newtheorem{proposition}{Proposition}[section]

\newtheorem{remark}{Remark}[section]

\def\oH{\buildrel\circ\over H}
\def\oH1{\buildrel\circ\over H\kern-.02in{}^1}

\def\l{\ell}

\begin{document}


\title{An inverse problem for the heat equation}

\author{
A.G. Ramm\\
 Mathematics Department, Kansas State University, \\
 Manhattan, KS 66506-2602, USA\\
ramm@math.ksu.edu\\
}

\date{}

\maketitle\thispagestyle{empty}

\begin{abstract}
Let $u_t = u_{xx} - q(x) u, 0 \leq x \leq 1$, $t>0$,
$u(0, t) = 0, u(1, t) = a(t), u(x,0) = 0$,
where $a(t)$ is a given function vanishing for $t>T$, $a(t) \not\equiv 0$,
$\int^T_0 a(t) dt < \infty$. Suppose one measures the flux
$u_x (0,t) := b_0 (t)$ for all $t>0$. Does this information determine
$q(x)$ uniquely? Do the measurements of the flux $u_x (1,t) := b(t)$
give more information about $q(x)$ than $b_0 (t)$ does?

The above questions are answered in this paper.

\end{abstract}


\section{Introduction}
Consider the heat transfer problem described by the equation
$$u_t = u_{xx} - q(x) u, \quad 0 \leq x \leq 1, \quad t>0,
  \eqno{(1.1)}$$
$$u(x,0) = 0, \eqno{(1.2)}$$
$$u(0,t) = 0, \quad u (1,t) = a(t), \eqno{(1.3)}$$
where $a(t)$ is the prescribed temperature, and $q(x)$ is a real-valued
integrable function. Assume that $a(t)$ is a pulse-type function, that is,
$$a(t) = 0 \hbox{\ for\ } t>T, \quad \int^T_0 a(t) dt < \infty,
  a(t) \not\equiv 0. \eqno{(1.4)}$$
In particular, one can choose $a(t)$ to be the delta-function 
$a(t)=\delta (t).$
Suppose one measures the flux at one of the ends of the rod, either measuring
$$u_x (1,t) := b(t), \eqno{(1.5)}$$
or
$$u_x (0,t) := b_0 (t). \eqno{(1.6)}$$

The questions that are answered in this paper are:

1) Does the knowledge of $a(t)$ and $b_0(t)$ for all $t>0$ determine
$q(x)$ uniquely?

2) Does the knowledge of $a(t)$ and $b(t)$ for all $t>0$ determine
$q(x)$ uniquely?

3) How does one calculate $q(x)$ given $a(t)$ and $b(t)$?

The answers we give are:

1) The knowledge of $a(t)$ and $b_0 (t)$ for all $t>0$ does not determine
$q(x)$ uniquely, in general. It does, if $q(x)$ is symmetric with respect
to the point $x = \frac{1}{2}$, that is, if
$q(x + \frac{1}{2}) = q(\frac{1}{2} - x)$, or if $q(x)$ is known on the
interval $[\frac{1}{2}, x]$.

2) The knowledge of $a(t)$ and $b(t)$ for all $t>0$ determines $q(x)$
uniquely.

3) An algorithm for computing $q(x)$ given $a(t)$ and $b(t)$ is given.

The answer to question 2) was given in \cite{R1} and earlier, under the
additional assumption $q(x) \geq 0$, in \cite{D}. The answer to
question 1) is new, as far as the author knows. An algorithm for computing
$q(x)$ is similar to the one described in \cite{R1}.

In section 2 the answers to questions 1) and 2) are given, 
and
an answer to question 3) is given in section 3.

\section{Answer to questions 1) and 2).}
Let us Laplace-transform (1.1)-(1.3), (1.5) and (1.6). If
$v := v(x,\lambda) := \int^\infty_0 u(x,t) e^{-\lambda t} dt$,
then
$$v^{\prime \prime} - q(x)v - \lambda v = 0, \quad 0 \leq x \leq 1
  \eqno{(2.1)}$$
$$v(0, \lambda) = 0, \quad v (1, \lambda) = A(\lambda), \eqno{(2.2)}$$
$$v_x (1, \lambda) = B(\lambda), \eqno{(2.3)}$$
$$v_x(0,\lambda) = B_0 (\lambda), \eqno{(2.4)}$$
where $A(\lambda)$, $B(\lambda)$ and $B_0(\lambda)$ are the Laplace
transforms of $a(t)$, $b(t)$ and $b_0 (t)$, respectively.

\begin{proposition} 
The data $\{A(\lambda), B(\lambda)\}$ known for a set of $\lambda >0$,
which has a finite limiting point, determines $q(x)$ uniquely.
\end{proposition}

\begin{proposition} 
The data $\{A(\lambda), B_0 (\lambda)\}$, known for all $\lambda >0$, does
not determine $q(x)$ uniquely, in general.

If $q(x)$ is known on the interval $\left[\frac{1}{2}, 1 \right]$
then $q(x)$ on the interval $\left[ 0, \frac{1}{2} \right]$ is uniquely
determined by the above data.

Also, if $q(\frac{1}{2} - x) = q(\frac{1}{2} + x)$, then $q(x)$ is
uniquely determined on the interval $[0,1]$ by the above data.
\end{proposition}

\begin{proof}[Proof of Proposition 2.1]

Let $\varphi (x,\nu)$ solve equation (2.1) with $\lambda = -\nu$ and satisfy
the condition
$$\varphi (0, \nu) = 0, \quad \varphi^\prime (0, \nu) = 1. \eqno{(2.5)}$$
The solution $\varphi(x,\nu)$ is an entire function of $\nu$ and of
$k=\nu^{1/2}=i \lambda^{1/2}$ (see \cite{L}, \cite{R2}).

Since $\varphi$ and $v$ satisfy the first condition (2.2), one has:
$$v(x, \lambda) = c(\lambda) \varphi(x, -\lambda), \eqno{(2.6)}$$
where $c(\lambda)$ does not depend on $x$. Thus
$$c(\lambda) \varphi(1, -\lambda) = A(\lambda), \quad c(\lambda)
  \varphi^\prime (1, - \lambda) = B(\lambda). \eqno{(2.7)}$$
Note that $v(x, \lambda)$ may be not defined for some $\lambda$, namely for
some $\lambda$, namely for $-\lambda = \lambda_j$, where $\lambda_j$ are
the eigenvalues of the problem
$$\l \psi_j := \psi_j^{\prime \prime} + q(x) \psi_j =
  \lambda_j \psi_j, \quad \psi_j (0) = \psi_j (1) = 0.
  \eqno{(2.8)}$$
Since $\lambda >0$, the condition $\lambda = -\lambda_j$ can be satisfies
only if $\lambda_j <0$. There are at most finitely many negative
eigenvalues of the selfadjoint Dirichlet operator
$\l = -\frac{d^2}{dx^2} + q(x)$ in $H := L^2[0,1]$. For the problem
(2.1)-(2.2) to be solvable, when $\lambda = -\lambda_j$ it is necessary and
sufficient that the appropriate orthogonaltiy conditions are satisfied.
Namely one finds
$$v(x, \lambda) = -\sum ^\infty_{j=1}
  \frac{A(\lambda) \psi_j^\prime (1)}{\lambda + \lambda_j} \psi_j (x).
  \eqno{(2.9)}$$

For this series to be defined at $\lambda = -\lambda_j >0$ it is necessary
and sufficient that $A(-\lambda_j) = 0$. Note that $\psi_j^\prime (1)\neq
0$ by the uniqueness of the solution to the Cauchy problem (see (2.8)).

Since we have assumed $a(t) = 0$ for $t>T$, the function $A(\lambda)$
is an entire function of $\lambda$ on the complex $\lambda$-plane. Therefore
$v(\lambda)$ is well-defined as a meromorphic function of the parameter
$\lambda$ with values in $H$. Note that problem (1.1)-(1.3) is always
solvable, but if the operator $\l$ has negative eigenvalues, then the
solution to (1.1)-(1.3) may grow exponentially as $t \to +\infty$.

From (2.7) one concludes
$$\frac{B(\lambda)}{A(\lambda)} =
  \frac{\varphi^\prime(1-\lambda)}{\varphi (1, -\lambda)},
  \eqno{(2.10)}$$
since $c(\lambda) \neq 0$. The zeros of the function
$$\varphi (1, \nu) = 0 \eqno{(2.11)}$$
are precisely the Dirichlet eigenvalues $\lambda_j$ of $\l$, while the zeros of the
function
$$\varphi^\prime (1, \nu) = 0 \eqno{(2.12)}$$
are precisely the eigenvalues of the problem
$$\l w_j = \mu_j w_j, \quad w_j (0)  = 0, \quad w^\prime_j (1) = 0.
  \eqno{(2.13)}$$

It is well known (see e.g. \cite{L}) that the knowledge of  $\{\lambda_j\}$
and $\{\mu_j\}$ for all $j$ determines $q(x)$ uniquely
because two spectra of $\l$
with the same homogeneous boundary condition at $x=0$ and two different
homogeneous boundary condition at $x=1$, determine $q(x)$ uniquely.

The zeros of
$\frac{B(\lambda)}{A(\lambda)} =
  \frac{\varphi'(1, \lambda)}{\varphi(1, \lambda)}$
are the numbers $\mu_j$ and only these numbers, while its poles are the
numbers
$\lambda_j$ and only these numbers.

Proposition 2.1 is proved. 

\end{proof}

\begin{remark} 
A different proof of Proposition 2.1, based on Property C for ODE, is given in
\cite{R1}.
\end{remark}

\begin{proof}[Proof of Proposition 2.2]

From (2.6) and (2.4) it follows that
$$c(\lambda) \varphi (1, -\lambda) = A(\lambda), \quad c(\lambda)
  \varphi^\prime (0, -\lambda) = B_0 (\lambda). \eqno{(2.14)}$$

Thus
$$\frac{B_0(\lambda)}{A(\lambda)} =
  \frac{\varphi^\prime (0, -\lambda)}{\varphi(1, -\lambda)} =
  \frac{1}{\varphi(1, -\lambda)} = \frac{1}{\varphi(1, \nu)}, \quad
  \nu := -\lambda. \eqno{(2.15)}$$
The poles of the function (2.15) are the eigenvalues $\lambda_j$, and this
is the only information one can get from (2.15).

The knowledge of one spectrum $\{\lambda_j\}$ of $\l$ determines, roughly
speaking, ``half of the potential": namely, if $q(x)$ is known on the
interval $\left[\frac{1}{2}, \l \right]$, then the data $\{\lambda_j\}$
known for all $j$ determine $q(x)$ on $\left[0, \frac{1}{2} \right]$
uniquely (see \cite{HL}, \cite{R1}, \cite{R3}). By the same reason if
$q(x + \frac{1}{2}) = q(\frac{1}{2} -x)$ then $q(x)$ is uniquely determined
on $[0,1]$ by the set $\{\lambda_j\}$ known for all $j$.

Proposition 2.2 is proved.
\end{proof}

The information in the data $a(t)$ and $b_0(t)$ is equivalent to
the information in the ratio   $\frac{B_0(\lambda)}{A(\lambda)}.$
This is especially clear if one takes $a(t)=\delta(t)$ because
in this case $A(\lambda)=1$ and $\frac{B_0(\lambda)}{A(\lambda)}=
\frac{1}{\varphi(1, \nu)},$ so that the information in the ratio
is given just by one function $\varphi(1, \nu).$

\begin{remark} 
In \cite{R1} and \cite{R3} a general uniqueness result is obtained which says
that if $q(x)$ is known on $[b,1]$, $0 < b <1,$
where $b$ is an arbitrary fixed number,
then the set $\{\lambda_{m(j)}\}$ determines $q(x)$ on $[0,b]$ uniquely
provided that  $\sigma \geq 2b$.
Here $\lambda_{m(j)}$ is an arbitrary subset of $\{\lambda_j\}$ such that
$m(j) = \frac{j}{\sigma} (1 + \varepsilon_j), \sum^\infty_{j=1}
  |\varepsilon_j| < \infty.$
So, if $m(j) = j$, then
$\sigma =1, \varepsilon_j = 0, b \leq \frac{1}{2}$.
For $b = \frac{1}{2}$ one gets the uniqueness result used in the proof of
Proposition 2.2 and obtained in \cite{HL}.
\end{remark}

\begin{remark} 
From our arguments it follows that extra data (1.6) yields, roughly
speaking, half of the information that data (1.5) yields, and 
therefore does not allow one to recover $q(x)$ uniquely.
\end{remark}

\section{An algorithm for computing $q(x)$} 
If $\{a(t), b(t)\}$ are our data, one takes the Laplace transform and gets
$\frac{B(\lambda)}{A(\lambda)}$. One calculates the zeros and poles of this
function and gets the numbers $\{\lambda_j\}$ and $\{\mu_j\}$. In the
literature (see \cite{L}) there is an algorithm for calculating the spectral
function $\rho(\lambda)$ of the operator $\l$ from the knowledge of
$\{\lambda_j\} \cup \{\mu_j\}$. If $\rho(\lambda)$ is found, then the
Gelfand-Levitan algorithm allows one to calculate $q(x)$ from $\rho(\lambda)$.
This algorithm is described in \cite{L}, \cite{R1}, \cite{R2}.

In this section we describe an algorithm
which is a version of the one described in [5], pp.297-299 (see 
[4], p. 57), which is quite different from the Gelfand-Levitan one
and may be numerically more stable.

Recall that
$$\varphi (x, \nu) = \varphi_0 (x, \nu) + \int^x_0 K(x,y) \varphi_0(y, \nu)
  dy, \quad \varphi_0 (x, \nu) := \frac{\sin (kx)}{k}, \quad
  k = \sqrt{\nu}, \eqno{(3.1)}$$
where $K(x,y)$ is the transformation kernel, and
$$q(x) = 2\frac{dK(x,x)}{dx}. \eqno{(3.2)}$$
Since $\varphi(y, \lambda_j) = 0$, one gets:
$$\int^1_0 K(1,y) \varphi_0 (y, \lambda_j) dy = - \varphi_0 (1, \lambda_j),
  \quad j = 1,2, \dots \eqno{(3.3)}$$
Since the set $\{\varphi_{0j}\} := \{\varphi_0 (y, \lambda_j)\}_{\forall
j}$,
forms a Riesz basis of $H=L^2[0,1]$, relations (3.3) allow one to
find $K(1,y)$. 

Recall that a basis $\{h_j\}$ of a Hilbert space $H$ is called a
Riesz basis if there is a linear bounded map $A$ and $A^{-1}$ is a linear
bounded
operator on $H$, such that $h_j = Af_j$, where $\{f_j\}$ is an orthonormal
basis of $H$ (see \cite{R8}, p. 148).

Numerically one may look for $K(1,y)$ of the form
$$K(1,y) = \sum^J_{j=1} c_j \varphi_0 (y, \lambda_j), \eqno{(3.4)}$$
substitute (3.4) into (3.3) and get a linear system for
$c_j, 1 \leq j \leq J$. Here $J$ is an arbitrary large positive integer. The
matrix of the linear system is the Gram matrix
$$(\varphi_{0j}, \varphi_{0m}) := \int^1_0 \varphi_{0j} (x)
  \overline{\varphi_{0m} (x)} dx, \quad \varphi_{0j} (x) :=
  \frac{\sin (k_j x)}{k_j}, k_j = \sqrt{\lambda_j},$$
which is not ill-conditioned since $\{\varphi_{0j}\}$ forms a Riesz
basis.

Differentiate (3.1) with respect to $x$ and set $\nu = \mu_j$ $x=1$ to get
$$0 = \varphi^\prime_0 (1, \mu_j) + K(1,1) \varphi_0 (1, \mu_j) +
  \int^1_0 K_x (1, y) \varphi_{0j} (y) dy, \quad j=1,2, \dots .
  \eqno{(3.5)}$$
These equations determine uniquely $K_x(1,y)$, since
$\varphi^\prime_0 (1, \mu_j), K(1,1)$ and $\varphi_0(1, \mu_j)$ are known
numbers. Thus we can compute $K(1,y)$ and $K_x(1,y)$, $0 \leq y \leq 1$,
from the data $\{a(t), b(t)\}$.

If $K(1,t)$ and $K_x(1,t)$ are known, then one can derive a Volterra
integral equation for the unknown
$U := \{q(x), K(x,y)\}$ (see \cite{R1}, p.56, and \cite{R4}).

In \cite{R4} it is proved that this equation can be solved by iterations, and
therefore $q(x)$ can be computed by an iterative process.

For convenience of the reader we write down the above integral equation
for $U:= \{q(x), K(x,y)\}$ and an iterative process for the solution of
this equation:
$$U = W(U) + h, \eqno{(3.6)}$$
where
$$W(U) := \left( \begin{aligned}
                -2 &\int^1_x q(s) K(s, 2x-s) ds \\
                \frac{1}{2} &\int_{D_{xy}} q(s) K(s,t) ds dt
                \end{aligned}
                \right)_, \eqno{(3.7)}$$
$D_{xy}$ is the region bounded by the straight lines $s=1, t-y = s-x$, and
$t-y = x-s$ on the $(s,t)$ plane,
$$h=\left(\stackrel{\hbox{\normalsize $f$}}{g} \right), \eqno{(3.8)}$$
$$f(x) := 2[K_y (1, 2x-1) + K_x (1, 2x-1)], \eqno{(3.9)}$$
$$g(x,y) = \frac{K(1,y+x-1) + K(1, y-x+1)}{2} - \frac{1}{2}
  \int^{y-x+1}_{y+x-1} K_s (1,t) dt. \eqno{(3.10)}$$
Note that $f$ and $g$ are computable from the data $K(1,x)$ and $K_x(1,x)$,
and (3.6) is a nonlinear Volterra-type equation for the unknown $q(x)$ and
$K(x,y)$.

It is proved in \cite{R1}, p 57 (and in \cite{R4}) that the iterative
process
$$U_{n+1} = W(U_n) + h, \quad U_0 = h \eqno{(3.11)}$$
converges (at a rate of geometric series) to
$U(x) = \left( \begin{aligned} q(x) \\ K(x,y) \end{aligned} \right)$.
The details concerning the functional space in the norm of which the
convergence hold are given in \cite{R1}.

\end{document}